\documentclass[10pt,twocolumn,prb,showpacs,superscriptaddress,floatfix,amsmath,amssymb,amsfonts,aps]{revtex4-1}
\usepackage{graphicx}
\usepackage{color}
\usepackage{multirow}
\newcommand{\compN}{Na$_2$CuP$_2$O$_7$}
\newcommand{\compL}{Li$_2$CuP$_2$O$_7$}
\newcommand{\AFM}{\text{AFM}}
\newcommand{\eff}{\text{eff}}
\begin{document}

\title{Magnetic model for $A_2$CuP$_2$O$_7$ ($A$\,=\,Na,\,Li) revisited:\\ one-dimensional versus two-dimensional behavior}

\author{S. Lebernegg}
\affiliation{Department of Materials Engineering and Physics, Universit\"{a}t Salzburg, 5020 Salzburg, Austria}

\author{A. A. Tsirlin}
\email{altsirlin@gmail.com}
\affiliation{Max Planck Institute for Chemical Physics of Solids, 01187 Dresden, Germany}

\author{O. Janson}
\affiliation{Max Planck Institute for Chemical Physics of Solids, 01187 Dresden, Germany}

\author{R. Nath}
\affiliation{Max Planck Institute for Chemical Physics of Solids, 01187 Dresden, Germany}
\affiliation{School of Physics, Indian Institute of Science Education and Research, Trivandrum-695016 Kerala, India}

\author{J.~Sichelschmidt}
\affiliation{Max Planck Institute for Chemical Physics of Solids, 01187 Dresden, Germany}

\author{Yu. Skourski}
\affiliation{Dresden High Magnetic Field Laboratory, Helmholtz-Zentrum
Dresden-Rossendorf, 01314 Dresden, Germany}

\author{G. Amthauer}
\affiliation{Department of Materials Engineering and Physics, Universit\"{a}t Salzburg, 5020 Salzburg, Austria}

\author{H. Rosner}
\email{rosner@cpfs.mpg.de}
\affiliation{Max Planck Institute for Chemical Physics of Solids, 01187 Dresden, Germany}


\begin{abstract}
We report magnetization measurements, full-potential band structure calculations, and microscopic modeling for the spin-1/2 Heisenberg magnets $A_2$CuP$_2$O$_7$ ($A$\,=\,Na,\,Li) involving complex Cu--O--O--Cu superexchange pathways. Based on a quantitative evaluation of the leading exchange integrals and the subsequent quantum Monte-Carlo simulations, we propose a quasi-one-dimensional magnetic model for both compounds, in contrast to earlier studies that conjectured on a two-dimensional scenario. The one-dimensional nature of $A_2$CuP$_2$O$_7$ is unambiguously verified by magnetization isotherms measured in fields up to 50~T. The saturation fields of about 40~T for both Li and Na compounds are in excellent agreement with the intrachain exchange $J_1\simeq 27$\,K extracted from the magnetic susceptibility data. The proposed magnetic structure entails spin chains with the dominating antiferromagnetic nearest-neighbor interaction $J_1$ and two inequivalent, nonfrustrated antiferromagnetic interchain couplings of about $0.01J_1$ each. A possible long-range magnetic ordering is discussed in comparison with the available experimental information.
\end{abstract}


\maketitle

\section{Introduction}
Quantum effects in magnetic systems have diverse implications in exotic ground states\cite{lee2008,giamarchi2008,balents2010} and interesting finite-temperature properties, such as spin transport\cite{[{For example: }][{}]znidaric2011} and multiferroic behavior.\cite{kagawa2010,zapf2010} While transition-metal compounds offer nearly endless opportunities for finding variable spin lattices with different types of exchange couplings, complete understanding of the magnetic phenomena requires detailed and accurate information on the spin lattice and the energies of individual interactions. It is particularly important to establish the dimensionality of the system and the presence of magnetic frustration. The latter may lead to degenerate ground states, while both are responsible for the strength of quantum fluctuations. 

In complex crystal structures, the precise evaluation of the spin lattice remains a challenging problem. Single experimental observations on thermodynamic properties or magnetic structure may lead to wrong conclusions regarding the nature of the spin lattice, as in the alternating-chain compound (VO)$_2$P$_2$O$_7$ initially considered as a spin ladder,\cite{johnston1987,*garrett1997} and the frustrated-spin-chain system LiCu$_2$O$_2$ that was also ascribed to spin ladders.\cite{DFT_LiCu2O2,comment2005,*reply2005,*INS_LiCu2O2} Moreover, even the correct topology of the spin lattice does not guarantee the precise evaluation of the frustration regime: see, for example, the studies of Li$_2$VOXO$_4$ with X = Si, Ge.\cite{melzi2000,*melzi2001,rosner2002,*rosner2003,bombardi2004} Generally, only a complete survey of the magnetic excitation spectrum or a comprehensive investigation of the ground state, thermodynamics, and electronic structure give unambiguous information on the underlying spin model.

A particularly deceptive situation concerns the interpretation of the low magnetic ordering temperature $T_N$, as compared to the effective energy scale of exchange couplings, which is typically measured by the Curie-Weiss temperature $\theta$. If $T_N\ll|\theta|$, long-range magnetic order is impeded by quantum fluctuations, but it is impossible to decide \emph{a priori} whether this effect originates from the low dimensionality, from the frustration, or from a combination of both. For example, a quasi-two-dimensional (2D) non-frustrated system should feature $T_N/|\theta|\geq 0.2$,\cite{yasuda2005} while any lower value of $T_N$ indicates sizable frustration. However, in a quasi-one-dimensional (1D) system $T_N/|\theta|$ may be as low as 0.01 even without frustration.\cite{yasuda2005} A recent example of such an ambiguous situation is (NO)Cu(NO$_3)_3$ that was initially understood as a strongly frustrated quasi-2D system according to the magnetic susceptibility data and supposedly low $T_N$.\cite{volkova2010} Subsequent electronic structure calculations did not find any signatures of the frustration, and rather showed the pronounced quasi-one-dimensionality that is a plausible reason for the low $T_N$.\cite{janson2010} While \emph{ab initio} computational methods are an invaluable tool for studying complex materials, independent experimental evidence is indispensable. In the following, we show how both experimental and computational methods are separately used for evaluating the dimensionality of the system. After the dimensionality is established, simple criteria, such as the $T_N/|\theta|$ ratio, can be applied to analyze the frustration.

\begin{figure}[tbp]
\includegraphics[width=8.6cm]{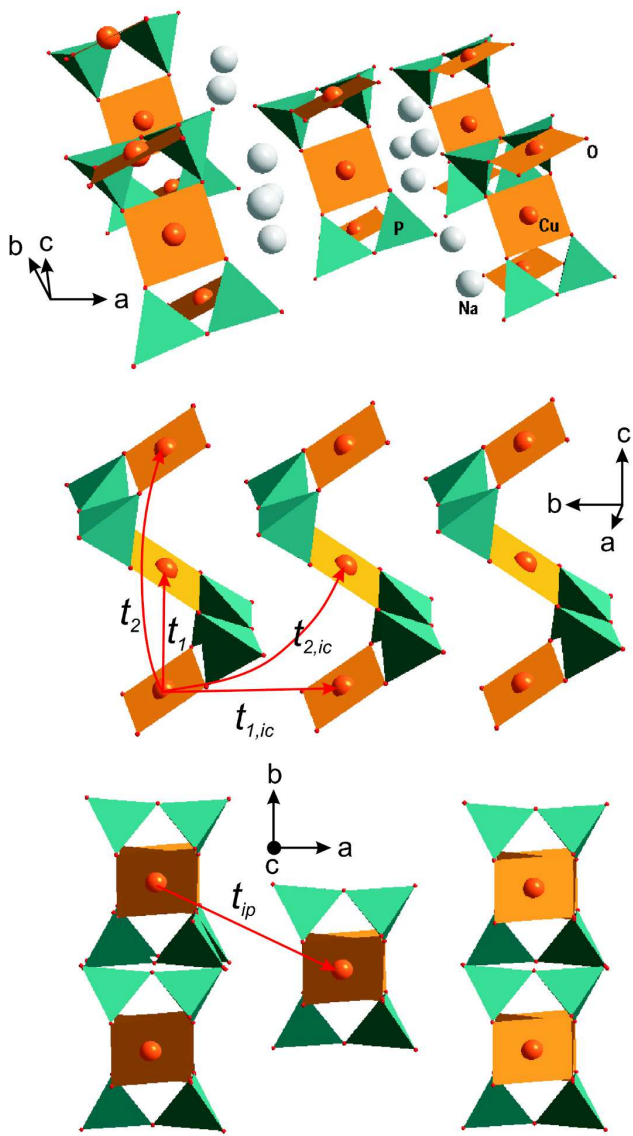}
\caption{\label{str}(Color online) Crystal structure of \compN~and the
leading magnetic coupling pathways. The indices $ic$ and $ip$ denote
interchain and interplane transfer, respectively. Apart from substitutions 
of Na by Li, the structure of \compL~shows only minor differences.}
\end{figure}

We present our approach for the rather simple and non-frustrated, albeit controversial spin-$\frac12$ model compounds  $A_2$CuP$_2$O$_7$ ($A$ = Li, Na). Both systems are antiferromagnetic (AFM) insulators and reveal apparent 1D features of the crystal structure (Fig.~\ref{str} and Sec.~\ref{sec:structure}). However, the magnetic dimensionality is hard to decide \emph{a priori}, because both intrachain and interchain couplings involve complex Cu--O--O--Cu superexchange pathways that can not be assessed in terms of simple empirical considerations, such as the Goodenough-Kanamori-Anderson rules.\cite{gka1,gka2,gka3} The experimental data for Na$_2$CuP$_2$O$_7$ reported by Nath~\mbox{\textit{et al.}\cite{nmr_na2cup2o7}} did not fit to the 1D spin model, and showed better agreement with the 2D scenario. Later, Salunke~\textit{et al.}\cite{asa_na2cup2o7} performed electronic structure calculations and argued that the leading couplings run along the structural chains, but the interchain couplings are stronger than in other quasi-1D compounds (e.g., in the chemically related K$_2$CuP$_2$O$_7$ presented in Ref.~\onlinecite{k2cup2o7}), and might be responsible for 2D features of the susceptibility at low temperatures. Here, we resolve the controversy and unequivocally establish the magnetic model of $A_2$CuP$_2$O$_7$ by revising both experimental and computational results. We confirm the quasi-1D scenario in an independent and more accurate electronic structure calculation, compared to the earlier work by Salunke~\mbox{\textit{et al.}\cite{asa_na2cup2o7}} We revisit the susceptibility data and present the results of high-field magnetization measurements that unambiguously evidence the quasi-1D nature of Na$_2$CuP$_2$O$_7$ and of its Li analog \compL.

The paper is organized as follows. In Section~\ref{sec:methods}, the applied experimental and computational methods are presented. Afterwards, in the third Section, the crystal structures are described and compared with similar compounds. In Section~\ref{sec:results}, the results of our theoretical investigations and the experimental data are discussed.  Finally, the conclusions and a short outlook are given in Section~\ref{sec:summary}.

\section{Methods}
\label{sec:methods}
Bluish-colored powder samples of \compN\ and \compL\ were prepared through solid-state reaction technique by mixing NaH$_{2}$PO$_{4}\cdot$H$_{2}$O (99.999\% pure) or LiH$_{2}$PO$_{4}\cdot$H$_{2}$O (99.99\% pure) and CuO (99.99\% pure) in appropriate molar ratios. The stoichiometric mixtures were fired at 800~$^{\circ}$C and 750~$^{\circ}$C for 60 hours each, with one intermediate grinding and pelletization to achieve single-phase \compN\ and \compL\ compounds, respectively. The purity of the samples was confirmed by x-ray diffraction (Huber G670 camera, CuK$_{\alpha1}$ radiation, ImagePlate detector, $2\theta=3-100^{\circ}$ angle range). 

Magnetic susceptibility was measured with the commercial MPMS SQUID magnetometer in the temperature range $2-380$~K in applied fields up to 5~T. High-field data were collected at a constant temperature of 1.5~K using pulsed magnet installed at the Dresden High Magnetic Field Laboratory. Details of the experimental procedure can be found elsewhere.\cite{tsirlin2009} 

Electron spin resonance (ESR) spectra were measured at room temperature using the $X$-band frequency of 9.4~GHz. The spectra were fitted as powder average of several Lorentzian lines reflecting different components of the anisotropic $g$-tensor.

The electronic structure was calculated using the density functional theory (DFT)-based full-potential local-orbital code (FPLO) version 9.00-34.\cite{fplo} For the scalar-relativistic calculations within the local density approximation (LDA), the
Perdew-Wang parameterization of the exchange correlation potential\cite{pw92} was used together with a well converged k-mesh of 10$\times$10$\times$10 points. The strong electron correlations, only poorly described in LDA, were considered: ($i$) by mapping the magnetically active (partially filled) bands first onto a tight-binding (TB) model
\begin{equation}
\hat{H}_{TB}=\sum_{i}\varepsilon_{i}+\sum_{\left\langle ij
\right\rangle\sigma}t_{ij}\left(\hat{c}^{\dagger}_{i,\sigma}\hat{c}_{j,\sigma}+H.c.\right)
\end{equation}
and in a second step onto a single-band Hubbard model $\hat{H}=\hat{H}_{TB}+U_{\text{eff}}\sum_{i}\hat{n}_{i\uparrow}\hat{n}_{i\downarrow}$.
In the strongly correlated limit, $U_{\text{eff}}\gg t_{ij}$, and for half-filling, well justified for undoped cuprates, the low-energy sector of the Hubbard model may further be mapped onto a Heisenberg model
\begin{equation}
\hat{H}=\sum_{\left\langle ij\right\rangle}J_{ij}\hat{S_{i}}\hat{S_{j}},
\end{equation}
yielding the AFM contribution to the superexchange coupling constant $J$ in second order as $J^{\text{AFM}}_{ij}=4t^2_{ij}/U_{\text{eff}}$. 
($ii$) Alternatively, the effects of correlation were considered in a mean-field way within the local spin density approximation (LSDA)+$U$ method ($U_{d}$\,=\,$7\pm0.5$\,eV, $J_{d}$\,=\,1\,eV).\cite{dioptas,janson2010,janson2011} For the double-counting correction, the around-mean-field approach\cite{dcc} was applied.\cite{tsirlin2010} The magnetic coupling constants $J_{ij}$ were obtained from energy differences between several collinear spin configurations.
 
Quantum Monte-Carlo (QMC) simulations were performed using the \texttt{looper}\cite{loop} and \verb|dirloop_sse| (directed loop in the stochastic series expansion representation)\cite{dirloop} algorithms of the software package \texttt{ALPS}.\cite{ALPS} The magnetization was simulated for the $N=400$-site chain with periodic boundary conditions. To evaluate the ordering temperature $T_N$, we considered the full three-dimensional (3D) spin lattice entailing interchain couplings, and calculated the Binder ratio\cite{binder} $B=\langle{M_s^4}\rangle/\langle{M_s^2}\rangle^2$ for the staggered magnetization $M_s$ on the cluster size $N$. We performed a series of simulations starting with a $N$\,=\,640-sites cluster and consequently increasing it up to $N$\,=\,40960 sites. In the temperature range $0.1\le\,T/J_1\,\le\,18.0$, we used periodic boundary conditions, 20~000 sweeps for thermalization and 200~000 sweeps after thermalization.  The resulting statistical errors (below 0.1\,\%) are negligible compared to the experimental error bars. Magnetic susceptibility of a Heisenberg model on a square lattice was computed on the $N$\,=\,20$\times$20 finite lattice, using 30 000 sweeps for thermalization and 300 000 sweeps after thermalization.
 
\section{Crystal structure}
\label{sec:structure} 
Na$_{2}$CuP$_{2}$O$_{7}$ crystallizes in two dissimilar monoclinic structures.\cite{structure_na2cup2o7} In the following, we focus on the \mbox{$\beta$-polymorph} that is isostructural to Li$_2$CuP$_2$O$_7$. \mbox{$\beta$-Na$_2$CuP$_2$O$_7$}, further referred as Na$_2$CuP$_2$O$_7$, has the space group $C2/c$ with the lattice parameters $a=14.728$\,\AA, $b=5.698$\,\AA, $c=8.067$\,\AA, and $\beta=115.15^{\circ}$.\cite{structure_na2cup2o7} The crystal structure entails chains stretched along the $c$ direction. Each chain consists of CuO$_{4}$ plaquettes linked via two corner-sharing PO$_{4}$ tetrahedra. Neighboring plaquettes are tilted toward each other by an angle of about $70^{\circ}$ (Fig.~\ref{str}). Compared to a planar arrangement of the plaquettes in other quasi-1D Cu$^{+2}$ phosphates, such as K$_2$CuP$_2$O$_7$ exhibiting a pronounced quasi-1D magnetic behavior,\cite{k2cup2o7} the tilting could enhance the interchain couplings and might be responsible for the proclaimed quasi-2D magnetism.\cite{nmr_na2cup2o7,asa_na2cup2o7} On the other hand, CuSe$_2$O$_5$ with a similar tilting angle as in \compN~was clearly shown to be of 1D type,\cite{cuse2o5} although the distances between the $bc$-planes are significantly smaller and even the chains within the $bc$-plane are somewhat closer together than in \compN. Accordingly, it could be expected from structural considerations that \compN~is a quasi-1D system.
\begin{figure}[tbp]
\includegraphics[angle=270,width=8.3cm]{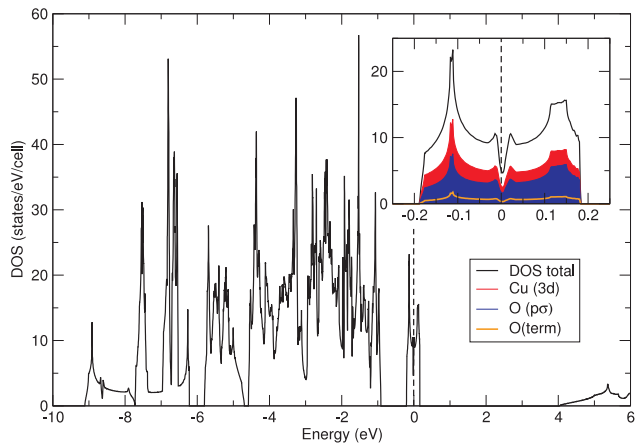}
\caption{\label{dost}(Color online) LDA electronic density of states and
partial density of states at the Fermi level (inset) for \compN~arising
from the two well separated bands (see Fig.~\ref{bandNL}). O($p\sigma$)
denotes the contribution from the $\sigma$-antibonding $2p$-orbitals of
the O atoms being part of the plaquettes. O(term) denotes the
contributions from O-atoms being not part of the plaquettes. The results
for \compL~are very similar (see Fig.~\ref{bandNL}).} \end{figure}
\begin{figure}[tbp]
\includegraphics[angle=270,width=8.3cm]{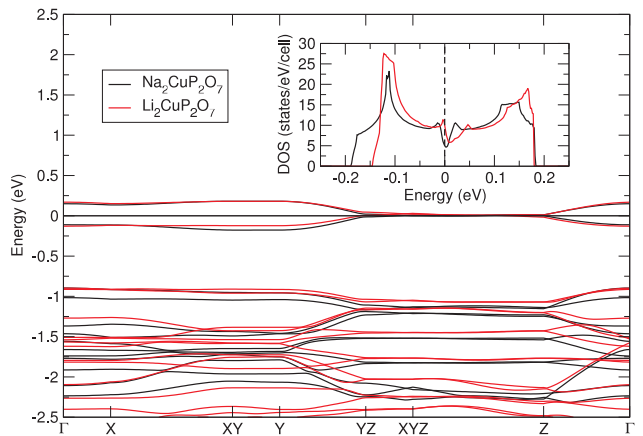}
\caption{\label{bandNL}(Color online) Comparison between the LDA band structures of \compN~and \compL. The inset shows the DOS of both compounds at the Fermi level.} 
\end{figure}

\compL~is isostructural to $\beta$-\compN~with the unit cell parameters $a=15.336$\,\AA, $b=4.8733$\,\AA, $c=8.6259$\,\AA, and $\beta=114.79^{\circ}$.\cite{listr,spirlet1993} The tilting angle is about $90^{\circ}$, leading to shorter distances between
the chains within the $bc$-plane, where in particular the $t_{1,ic}$ path is significantly shorter. The distances between the planes are only slightly smaller than in \compN. Since in these compounds the plaquettes are linked via the
phosphate tetrahedra, the empirical Goodenough-Kanamori-Anderson rules\cite{gka1,gka2,gka3} cannot be applied for describing the NN superexchange as a function of the tilting angle, and an elaborate analysis of the electronic structure is required.

\section{Results and Discussion}
\label{sec:results}
\subsection{DFT calculations}

We start with the computational analysis of $A_2$CuP$_2$O$_7$. The DFT calculations of the band structure and the electronic density of states (DOS) within the LDA yield a valence band width of about 9\,eV and 9.5\,eV (Fig.~\ref{dost}) for \compN~and \compL, respectively, typical for cuprates. The band structures of the two compounds are very similar and show well separated bands crossing the Fermi level (Fig.~\ref{bandNL}) somewhat narrower in the case of \compL. Typical for cuprates, these bands are formed by $\sigma$-antibonding linear combinations of Cu($3d_{x^2-y^2}$) and O($2p$) orbitals (Fig.~\ref{dost}, inset). The orbitals are denoted with respect to a local coordinate system, where for each plaquette one of the Cu-O bonds and the direction perpendicular to the plaquette are chosen as $x$- and $z$-axes, respectively.

\begin{table}[tbp]
\begin{ruledtabular}
\caption{\label{T_tJ} Transfer integrals ($t_i$) and AFM contributions to the exchange integrals ($J^{\text{AFM}}_i=4t^2_i/U_{\text{eff}}$) for \compN~and \compL, as obtained from full-potential LDA calculations. The hoppings by Salunke~\textit{et al.}\cite{asa_na2cup2o7} were calculated for Na$_2$CuP$_2$O$_7$ within the ASA (see text for details).}
\begin{tabular}{c c c c c c}
(meV)& $t_1$ & $t_2$ & $t_{1,ic}$ & $t_{2,ic}$ & $t_{ip}$  \\
\compN  & 75.4 & 0.8 & 0.3 & $-6.9$ & 6.9 \\ 
  Salunke~\textit{et al.}\cite{asa_na2cup2o7} & 55.8 & 1.4 & 1.4 & 5.4 & 4.1  \\
\compL & 71.5 & 4.3 & $-6.8$ & $-2.3$ & 8.2 \\
& & & & &  \\
(K) & $J^{\text{AFM}}_{1}$ & $J^{\text{AFM}}_{2}$ & $J^{\text{AFM}}_{1,ic}$ & $J^{\text{AFM}}_{2,ic}$ & $J^{\text{AFM}}_{ip}$ \\
\compN  & 58.7 & 0 & 0 & 0.5 & 0.5 \\ 
\compL & 52.7 & 0.2 & 0.5 & 0.1 & 0.7 \\
\end{tabular}
\end{ruledtabular}
\end{table}
Since the two bands at the Fermi level corresponding to the two Cu atoms per unit cell are well separated from the rest of the valence band, the magnetic properties can be described using an effective one-orbital model. Although LDA yields a wrong metallic ground state due to the poor description of strong electronic correlations, the LDA bands can be used as an input for an effective Hubbard model that introduces the missing correlations and yields the AFM part of the exchange, as described in Sec.~\ref{sec:methods}. A crucial parameter in this computational scheme is the LDA bandwidth $W$ determined by individual hoppings $t_{i}$. The bandwidth may be affected by the approximations introduced in the DFT calculation. Particularly, the atomic spheres approximation (ASA) for the potential is known to underestimate the bandwidth in Cu$^{+2}$ compounds.\cite{comment_ASA} Since the previous computational work by Salunke~\textit{et al.}\cite{asa_na2cup2o7} was based on the ASA calculation, the insufficient accuracy of the band structure could lead to a sizable error in the hoppings and exchange couplings, thus hindering the reliable evaluation of the dimensionality. 

\begin{figure}[tbp]
\includegraphics[angle=270,width=8.3cm]{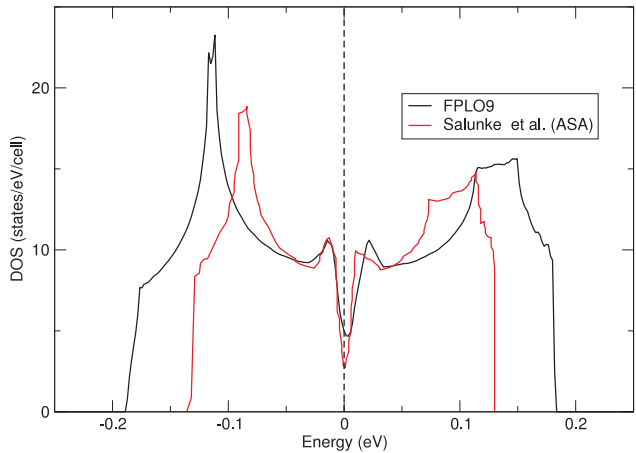}
\caption{\label{fpasa}(Color online) Comparison between the density of
states at the Fermi level calculated using a full-potential code (this
study) and an ASA code (Ref.~\citenum{asa_na2cup2o7}).} \end{figure}

\begin{figure}[tbp]
\includegraphics[angle=270,width=8.3cm]{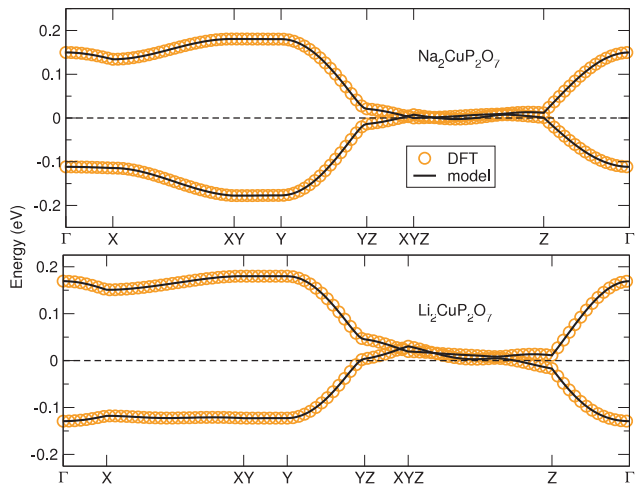}
\caption{\label{wf}(Color online) The LDA bands at the Fermi level and
the band dispersions calculated with a tight-binding model for \compN~(upper panel)
and \compL~(lower panel).} \end{figure}

To avoid the problems of the ASA, we performed full-potential LDA calculations that produce a highly accurate estimate of the LDA bandwidth and individual hoppings.\cite{comment_ASA} The hoppings are obtained from Wannier functions with the Cu $3d_{x^2-y^2}$ orbital character,\cite{FPLO_WF} and perfectly reproduce the LDA bands, as demonstrated in Fig.~\ref{wf}. In Table~\ref{T_tJ}, we compare our results for both \compL\ and \compN\ with the previous ASA data by Salunke~\textit{et al.}\cite{asa_na2cup2o7} for \compN. In the ASA calculations, the bandwidth $W$ (Fig.~\ref{fpasa}) and the intrachain hopping $t_1\simeq W/4$ are indeed underestimated for about 25~\%, which is similar to our previous comparative study of the ASA and full-potential calculations for Sr$_2$Cu(PO$_4)_2$.\cite{comment_ASA} 

Applying the effective on-site Coulomb repulsion $U_{\text{eff}}=4.5$~eV,\cite{k2cup2o7,sr2cupo42,tsirlin2010} we evaluate AFM contributions to individual exchange couplings using the second-order expression $J_i^{\AFM}=4t_i^2/U_{\eff}$. (Table~\ref{T_tJ}).%
\footnote{$U_{\text{eff}}$ is an effective quantity that accounts for the charge screening effects. Since this quantity can not be measured directly, the optimal value is defined empirically, by extensive comparisons between calculated and experimental exchange couplings for a range of systems with similar features of the crystal structure. Particularly, the estimate of $U_{\eff}=4.5$~eV has been confirmed in our recent studies of the related compounds Sr$_2$Cu(PO$_4)_2$ and Ba$_2$Cu(PO$_4)_2$ (Ref.~\onlinecite{sr2cupo42}), K$_2$CuP$_2$O$_7$ (Ref.~\onlinecite{k2cup2o7}), and \mbox{$\beta$-Cu$_2$V$_2$O$_7$} (Ref.~\onlinecite{tsirlin2010}). Note also that $U_{\eff}$ introduces a simple scaling of $J_i^{\AFM}$, while the ratios of the exchange integrals are solely determined by the hopping parameters. Therefore, the specific choice of the $U_{\eff}$ value does not alter our conclusions on the dimensionality of the spin system.} Since the spin lattice does not contain triangular units,\cite{bulaevskii2008} the higher-order corrections\cite{stein1997,*reischl2004,delannoy2005} are proportional to $t_i^4/U_{\eff}^3$ and rather small compared to both intrachain and interchain couplings $J_i^{\AFM}$. For example, the fourth-order correction to the bilinear exchange is $-24t_1^4/U_{\eff}\simeq -0.1$~K and the leading four-spin term is $80t_1^2t_{\alpha}^2/U_{\eff}^3\simeq 0.03$~K, where $t_{\alpha}=t_{2,ic},t_{ip}$.\cite{delannoy2005} Both fourth-order terms are comparable to or even smaller than typical dipolar interactions in a spin-$\frac12$ compound.\cite{tong2010}

While individual hoppings derived from the full-potential and ASA calculations are rather different (note, e.g., the different signs of $t_{2,ic}$), the resulting microscopic scenario is very similar. The interchain couplings $J_{2,ic}$ and $J_{ip}$ are about two orders of magnitude smaller than the intrachain coupling $J_1$. Although Salunke~\textit{et al.}\cite{asa_na2cup2o7} claim that ``the interchain interactions in \compN\ can not be neglected'', our estimates of $J_{2,ic}^{\AFM}$ and $J_{ip}^{\AFM}$ suggest a pronounced 1D character of $A_2$CuP$_2$O$_7$. In fact, \compN\ features comparable couplings along the two interchain directions ($J_{2,ic}^{\AFM}\simeq J_{ip}^{\AFM}$), and should be considered either as a quais-1D or as a three-dimensional (3D) system, in contrast to the 2D behavior proposed in the experimental study.\cite{nmr_na2cup2o7}

While the discrepancy between the computational and experimental results is actually resolved on the experimental side (Sec.~\ref{sec:exp}), we also check for possible shortcomings of our calculations, and supplement our model estimate of $J_i^{\AFM}$ with the full exchange couplings $J_i$ obtained from LSDA+$U$ calculations. Such full exchange couplings were not evaluated by Salunke~\textit{et al.}\cite{asa_na2cup2o7} The FM contributions considered in the LSDA+$U$ calculations reduce the NN coupling constant $J_1$ to $35\pm4$\,K and $34\pm4$\,K for \compN~and \compL, respectively (compare to $J_1^{\AFM}$ in Table~\ref{T_tJ}). Interchain and interplane couplings were estimated to be smaller than 0.2\,K. Thus, the LSDA+$U$ results further support the 1D scenario. Similar ratios of the interchain ($J_{ic}/J_1$) and the interplane ($J_{ip}/J_1$) couplings were found in several other quasi-1D systems.\cite{cuse2o5,sr2cupo42} In conclusion, unlike in previous studies\cite{asa_na2cup2o7} we find no indications for the 2D behavior neither for \compN~nor for \compL~based on electronic structure calculations.

\subsection{Experimental data}
\label{sec:exp}
After establishing the quasi-1D microscopic scenario, we re-consider the experimental data to check whether the previous conclusion on the 2D magnetic behavior\cite{nmr_na2cup2o7} was well-justified. The conjecture about the 2D magnetism made by Nath~\emph{et~al.}\cite{nmr_na2cup2o7} was essentially based on the better agreement with the experimental data (nuclear-magnetic-resonance shift as a measure of intrinsic magnetic susceptibility), in comparison to a 1D model. However, our precise QMC simulations for both 1D and 2D models challenge this interpretation. 

\begin{figure}[tbp]
\includegraphics[angle=270,width=8.3cm]{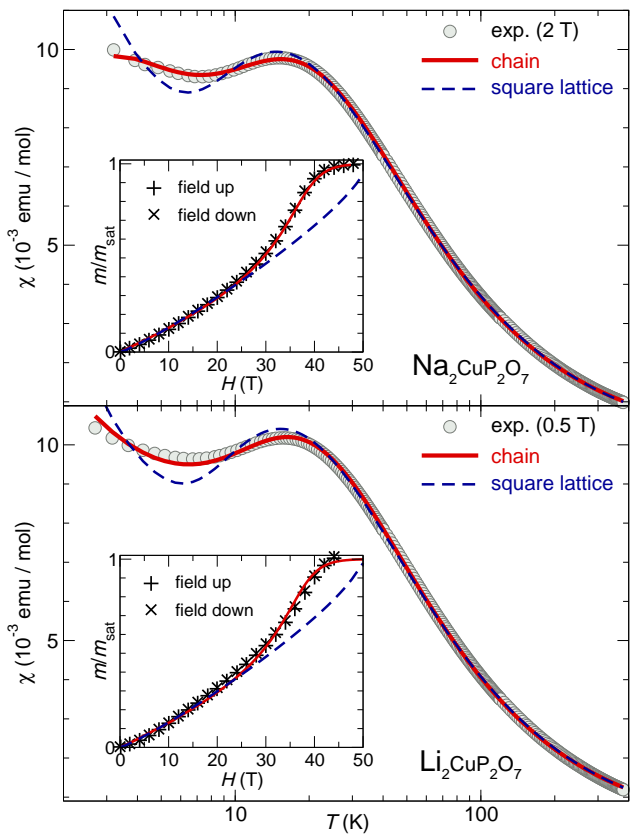}
\caption{\label{F_oj_chiT}(Color online) 
Fits of the magnetic susceptibility of \compN\ (top) and \compL\ (bottom). Magnetic susceptibility was simulated using QMC for a Heisenberg nearest-neighbor one-dimensional (chain) and two-dimensional (square lattice) models. Insets: simulated magnetization curves on top of the experimental high-field magnetization data. For the scaling, the parameters from the fits to $\chi(T)$ were used (see text for details).\cite{footnote}
}
\end{figure}

Magnetic susceptibility ($\chi$) of the Li and Na compounds is very similar and shows the maximum around 16~K, as typical for low-dimensional magnets (Fig.~\ref{F_oj_chiT}). The increase in $\chi$ below 7~K is likely related to the Curie-like contribution of paramagnetic impurities. We fit the data assuming simplest 1D and 2D coupling topologies: a NN chain and a square lattice.  Reduced magnetic susceptibility $\chi^*$ was simulated using QMC for large finite lattices of $N$\,=\,400 and $N$\,=\,20$\times$20 spins, for the chain and the square lattice model, respectively. The simulated
$\chi^{*}(T/[k_B\,J_1])$ dependency was fitted to the experimental curves using the expression:
\begin{equation}
\label{E-chiT-fit}
\chi(T)=
\frac{N_A\,g^2\,{\mu}_{B}^2}{k_B\,J_1}\cdot\chi^{*}\biggl(\frac{T}{k_B\,J_1}\biggl)
+ \frac{C_{\text{imp}}}{T} + \chi_0,
\end{equation}
where the fitted parameters are the exchange coupling $J_1$, the Land\'e factor $g$, the extrinsic (impurity and defect) paramagnetic contribution $C_{\text{imp}}/T$, and the temperature-independent term $\chi_0$ that accounts for the core diamagnetism and van Vleck paramagnetism. 

The resulting fits are shown in Fig.~\ref{F_oj_chiT}, with the fitting parameters listed in Table~\ref{T_fit}. For both \compN\ and \compL, the square-lattice model apparently fails to describe $\chi(T)$ below 10\,K. In contrast, the NN chain model provides a much better description of the data down to 2\,K. This contradicts the previous conclusion by Nath~\textit{et al.},\cite{nmr_na2cup2o7} although our susceptibility data for \compN\ are very similar to those previously reported. The discrepancy originates from different model expressions used in our analysis. Nath~\textit{et al.}\cite{nmr_na2cup2o7} utilize the high-temperature expressions that are valid down to $T\simeq J_1/2\simeq 14$~K only. Above 14~K, the 1D and 2D fits nearly coincide, hence neither model can be chosen unambiguously. The QMC simulations enable the precise evaluation of the susceptibility down to the lowest temperature of 2~K accessed in the present experiment. The low-temperature susceptibility data clearly favor the 1D model, while rule out the 2D description of $A_2$CuP$_2$O$_7$. Note that the experimental estimates of $J_1\simeq 27$\,K within the 1D model are in reasonable agreement with $J_1\simeq 35$\,K computed by LSDA+$U$.

Four adjustable parameters of the susceptibility fit might lead to an ambiguous solution. Therefore, we verified the fitted $g$-values by an ESR measurement. The room-temperature ESR spectrum of Li$_2$CuP$_2$O$_7$ was well described by two components of the $g$-tensor, $g_{\|}=2.064(2)$ and $g_{\perp}=2.198(2)$. In case of Na$_2$CuP$_2$O$_7$, three components were required: $g_{\alpha}=2.217(5)$, $g_{\beta}=2.152(3)$, and $g_{\gamma}=2.085(6)$. Both sets of the $g$-values result in the powder-averaged $\bar g\simeq 2.15(1)$, in good agreement with the fitted values of 2.12 and 2.19 for Li$_2$CuP$_2$O$_7$ and Na$_2$CuP$_2$O$_7$, respectively (see the fits for the chain model in Table~\ref{T_fit}).\footnote{%
Note that the fitted $g$-values are basically scale factors for the susceptibility ($\chi\sim g^2$). Therefore, they include the uncertainties related to the sample mass (about 1~\%). Additionally, the impurity contribution reduces the amount of the magnetic phase and also modifies the fitted $g$-value.} The three $g$-values resolvable in the ESR spectrum of Na$_2$CuP$_2$O$_7$ might be related to the distorted CuO$_4$ plaquette (Cu--O bond lengths of $1.91-1.96$~\r A) compared to the nearly regular CuO$_4$ plaquette in Li$_2$CuP$_2$O$_7$ (four Cu--O bonds of 1.93~\r A). 

Although the susceptibility fits are already a good evidence for the quasi-1D magnetism, one might still argue that the 1D model is solely chosen based on the low-temperature data that contain sizable impurity contributions. Moreover, the ESR results do not allow to discriminate between the marginally different $g$-values obtained in the chain and square-lattice fits (Table~\ref{T_fit}). To underpin our conclusions, we measured magnetization isotherms up to 50~T, and observed the magnetic saturation of the $A_2$CuP$_2$O$_7$ compounds. The saturation field ($H_s$) is an independent measure of the exchange couplings. The comparison between $H_s$ from the magnetization curve and $J_1$ from the susceptibility fit is a simple and efficient test for the validity of a spin model.\cite{nath2008} Experimental data (insets to Fig.~\ref{F_oj_chiT}) show that the magnetization of both Li and Na compounds saturates above 40~T. The upward curvature below $H_s$ is a typical feature of low-dimensional spin systems, and is generally ascribed to quantum effects.\cite{tsirlin2009}

\begin{table}[tbp]
\begin{ruledtabular}
\caption{\label{T_fit} 
Fitting parameters for the 1D and 2D description of the magnetic susceptibility of $A_2$CuP$_2$O$_7$: the intrachain exchange coupling $J_1$, the $g$-value, the temperature-independent contribution $\chi_0$, and the Curie-like paramagnetic impurity contribution $C_{\text{imp}}$.
}
\begin{tabular}{l@{\hspace{2em}}c@{\hspace{1em}}c cc c}
model & $J_1$ & $g$ & $\chi_0$            & $C_{\text{imp}}$&  \\
      & (K)   &     & ($10^{-4}$~emu/mol) & (emu~K/mol)        \\\hline
\compN & & & & \\ 
chain          &  27.0 & 2.12 & $-8.2$ & 0.011 \\
square lattice &  18.6 & 2.17 & $15.6$ & 0.020 \\\hline
\compL & & & & \\ 
chain          &  27.5 & 2.19 & 8.6 & 0.008 & \\
square lattice &  18.8 & 2.23 & 2.9 & 0.018 & \\
\end{tabular}
\end{ruledtabular}
\end{table}
To enable the proper comparison between the experimental and simulated magnetization, we subtract the paramagnetic impurity contribution according to $C_{\text{imp}}$ obtained from the susceptibility fit ($2-5$\% of the paramagnetic impurity depending on the model and compound). In Fig.~\ref{F_oj_chiT}, we juxtapose the experimental data and the simulations%
\footnote{The magnetization curves shown in Fig.~\ref{F_oj_chiT} are simulated for $T/J_1=0.1$, i.e., $T\simeq 2.7$~K, which is somewhat larger than the experimental temperature of 1.5~K. The larger temperature was necessary to better fit the curve around $H_s$, because the experimentally observed bend at $H_s$ is slightly broader than expected at 1.5~K. This effect may be ascribed to the sample heating induced by the magnetocaloric effect, or to the weak anisotropy. The anisotropy is evidenced by the multiple $g$-values observed in the ESR experiments. Therefore, the saturation field depends on the orientation of the crystallites with respect to the magnetic field, and a range of saturation fields is observed in the powder experiment. By contrast, residual interchain couplings only shift the saturation field without affecting the shape of the saturation anomaly.}
for both 1D and 2D models using the parameters from Table~\ref{T_fit}, as derived from the susceptibility measurements. The excellent agreement between the 1D scenario and the magnetization data verifies our suggestion on the quasi-1D model, while the 2D model apparently fails to reproduce the high-field data. Note that our conclusion is robust with respect to the impurity contribution, because the deviations from the 2D model are observed at high fields, where the paramagnetic contribution is saturatued and field-independent.

We have shown that the high-field magnetization data are an efficient tool for distinguishing between different spin models and even determining the dimensionality of the system. It is instructive to consider how the magnetization curve resolves the ambiguity of the susceptibility fit. In simple spin models, the position of the susceptibility maximum uniquely determines the exchange coupling $J_1$. For example, $T_{\max}/J_1\simeq 0.64$ and $T_{\max}/J_1\simeq 0.95$ for the uniform chain\cite{kluemper2000} and square lattice, respectively. This explains the 30\% difference in the respective estimate of $J_1$ (Table~\ref{T_fit}). The saturation field is directly related to the energies of different magnetic states, according to $H_s=2J_1k_B/(g\mu_B)$ and $H_s=4J_1k_B/(g\mu_B)$ for the 1D and 2D cases, respectively. Therefore, the same saturation field $H_s$ leads to the 50\% difference in~$J_1$. 

The dissimilar dependences of $T_{\max}$ and $H_s$ on $J_1$ provide a key to discriminate between the 1D and 2D magnetism using the combination of susceptibility and high-field magnetization data. The sizable difference between the saturation fields of the 1D and 2D models ensures that this method can be applied even to weak couplings on the order of 5\,K that produce about 1~T difference in $H_s$. Systems with larger couplings are, of course, easier to evaluate, although the couplings above 50\,K shift the saturation field above the limits of present-day facilities. Our simulations shown in the insets to Fig.~\ref{F_oj_chiT} suggest that even the part of the curve right below $H_s$ (at $30-35$~T in the present case) may be sufficient to discriminate between the 1D and 2D models. However, the data at lower fields do not contain the necessary information, because below 30~T the magnetization curves for the 1D and 2D systems nearly coincide.

\subsection{Long-range ordering}
\label{sec:lro}
In Sec.~\ref{sec:exp}, we have shown that thermodynamic properties of $A_2$CuP$_2$O$_7$ above 2\,K are well reproduced by the purely 1D model of the uniform spin chain. The actual system is, however, only quasi-1D, because any chemical compound necessarily entails both intrachain and interchain couplings. In $A_2$CuP$_2$O$_7$, the non-frustrated interchain couplings $J_{ip}$ and $J_{1,ic}$ (Li$_2$CuP$_2$O$_7$) or $J_{2,ic}$ (Na$_2$CuP$_2$O$_7$) should manifest themselves at low temperatures and eventually lead to the long-range magnetic ordering. No definitive signatures of the magnetic ordering were reported by Nath~\textit{et al.},\cite{nmr_na2cup2o7} according to heat-capacity and nuclear magnetic resonance (NMR) measurements down to 2\,K, although the broadening of the NMR line could indicate the approach to the magnetic transition below 4\,K. Our susceptibility measurements down to 2\,K did not reveal any anomalies in both Li and Na compounds either, thus suggesting the lack of the long-range order down to 2\,K. 

\begin{figure}[tbp]
\includegraphics[angle=270,width=8.6cm]{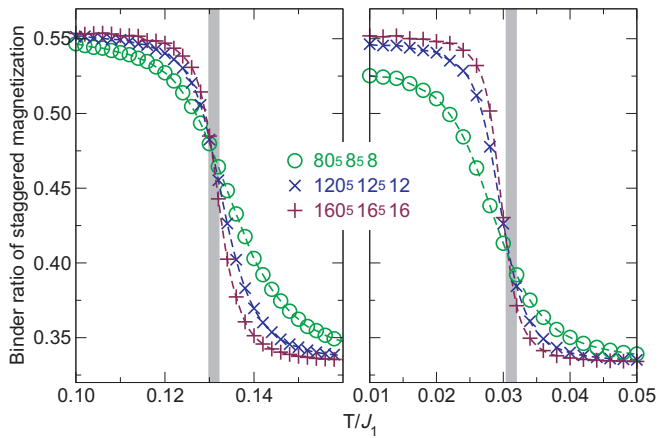}
\caption{\label{F_oj_brsm}(Color online) 
QMC simulations of the Binder ratio
$B(T)=\langle{}M_s^4\rangle/\langle{}M_s^2\rangle^2$ of
staggered magnetization $M_s$ for $J_{ic}/J_1$\,=\,0.05 (left panel) and
$J_{ic}/J_1$\,=\,0.01 (right panel). The finite lattice dimensions are given in
the legend. The ordering temperatures $T_N/J_1$ are
highlighted.}
\end{figure} 
The fact that the $A_2$CuP$_2$O$_7$ compounds do not order down to 2\,K ($T/J_1\simeq 0.08$) is an additional argument against the quasi-2D scenario, because the non-frustrated quasi-2D system orders at $T/J_1\geq 0.2$.\cite{yasuda2005} The low $T_N$ is, however, in excellent agreement with the quasi-1D scenario and indirectly supports our conclusions. To evaluate $T_N$, we considered the realistic 3D magnetic model of $A_2$CuP$_2$O$_7$ with the intrachain coupling $J_1$ as well as the interchain couplings $J_{2,ic}$ and $J_{ip}$ represented by a single effective interchain coupling $J_{ic}$. We utilize the scaling property of the Binder ratio ($B$) to be independent of the cluster size at $T_N$.\cite{binder} Therefore, $T_N$ is determined as the crossing point of $B(T)$ curves calculated for clusters of different size. The typical scaling is shown in Fig.~\ref{F_oj_brsm}, where we consider the cases of $J_{ic}/J_1=0.05$ and $J_{ic}/J_1=0.01$. For the former case, $T_N/J_1\simeq 0.13$ equals to 3.5~K and should be visible in the present experiments. Taking the weaker interchain coupling as the realistic estimate for \compN\ (Table~\ref{T_tJ}), we arrive at $T_N/J_1\simeq 0.032$, which is as low as 0.9\,K, well below the temperature range studied in our work. In \compL, a similar $T_N$ is expected, because $J_{1,ic}$ is comparable to $J_{2,ic}$ in the Na compound.

The anticipated $T_N/J_1\simeq 0.032$ in \compN\ is not the lowest transition temperature reported for quasi-1D magnetic systems. For example, Sr$_2$Cu(PO$_4)_2$ lacks the long-range magnetic order down to $T_N/J=4.5\times 10^{-4}$ (Ref.~\onlinecite{belik2005,nath2005}). This anomalously low $T_N$ is, however, facilitated by the frustrated nature of interchain couplings.\cite{sr2cupo42} In other spin-chain compounds,\cite{janson2010} the interchain couplings are often anisotropic, i.e., the interchain couplings in the plane are stronger than along the third direction. In contrast to all these perplexing scenarios, \compN\ reveals an unusually simple geometry of the interchain couplings with similar interactions along $a$ and $b$. This coupling regime conforms to the quasi-1D model typically considered in theoretical studies (e.g., Refs.~\onlinecite{yasuda2005,schulz1996,irkhin2000}). Therefore, it may be interesting to explore the magnetic transition in \compN\ experimentally. 

Experimental evaluation of $T_N$ and long-range magnetic order will, on one hand, verify theoretical results for weakly coupled spin chains, and on the other hand challenge our microscopic magnetic model. One efficient test is the magnetic structure that should feature antiparallel spins along $a$ and $c$, yet parallel spins along $b$ owing to the ``diagonal'' interchain coupling $J_{2,ic}$ (Fig.~\ref{str}). The anticipated propagation vector is $\mathbf k=0$, because the crystallographic unit cell contains two Cu sites along $a$ and $c$, while the order along $b$ is FM.

\section{Summary and outlook}
\label{sec:summary}
Motivated by suggestions of a quasi-2D behavior of \compN, exhibiting a crystal structure similar to well known quasi-1D-compounds, we have reinvestigated its microscopic model along with the previously unexplored \compL\ compound. To this end, magnetization measurements, full-potential DFT calculations, and quantum Monte Carlo simulations have been applied. We independently confirmed the anticipated quasi-1D scenario by the experimental data and by numerical estimates of individual exchange couplings. The computed intrachain couplings are in excellent agreement with the experiment. The previous conjecture on the quasi-2D magnetic behavior is ascribed to the incomplete analysis of the magnetization data. 

The microscopic model of \compN\ is a rare example of a ``regular'' quasi-1D system with similar couplings along the two interchain directions. Therefore, this compound is an excellent prototype material for the simplest model of weakly coupled spin chains that was widely studied theoretically. A further experimental study of the anticipated long-range ordering at $T_N\simeq 0.9$\,K may be insightful, as outlined in Sec.~\ref{sec:lro}. 

Another interesting aspect of the $A_2$CuP$_2$O$_7$ compounds are magnetostructural correlations for the superexchange in Cu$^{+2}$ phosphates and other transition-metal compounds containing polyanions. The intrachain couplings in \compL\ and \compN\ ($J_1\simeq 27$~K) are much smaller than in K$_2$CuP$_2$O$_7$ ($J_1\simeq 141$\,K)\cite{k2cup2o7} and Sr$_2$Cu(PO$_4)_2$ ($J_1\simeq 150$\,K)\cite{sr2cupo42} with similar chains of CuO$_4$ plaquettes bridged by PO$_4$ tetrahedra. The difference could be ascribed to the alteration of the chain geometry: both K$_2$CuP$_2$O$_7$ and Sr$_2$Cu(PO$_4)_2$ feature the coplanar plaquettes, while in $A_2$CuP$_2$O$_7$ the neighboring plaquettes form an angle of $70-90^{\circ}$ (Sec.~\ref{sec:structure}). The buckling impedes the overlap of the $p\sigma$ oxygen orbitals on the O--O edges of the PO$_4$ tetrahedra, thereby reducing the exchange. However, this simple explanation does not hold for CuSe$_2$O$_5$, where the buckling angle of $64^{\circ}$ resembles that of $A_2$CuP$_2$O$_7$, but the strong intrachain exchange of $J\simeq 157$~K rather reminds of the planar geometry in K$_2$CuP$_2$O$_7$.\cite{cuse2o5} A plausible explanation is the different role of the non-magnetic P$^{+5}$ and Se$^{+4}$ cations, and a further study of this matter for a broader compound family could be insightful.

\acknowledgments
We are grateful to Deepa Kasinathan for starting the computational work on \compN\ and fruitful discussions. We also acknowledge the experimental assistance of Yurii Prots and Horst Borrmann in x-ray diffraction measurements. The high-field magnetization experiments were supported by EuroMagNET II under the EC contract 228043. S.L. acknowledges the funding from the Austrian Fonds zur F\"orderung derwissenschaftlichen Forschung (FWF). A.T. was supported by Alexander von Humboldt Foundation. R.N. was funded by MPG-DST (Max Planck Gesellschaft, Germany -- Department of Science and Technology, India) fellowship.


\begin{thebibliography}{60}
\bibitem [{foo()}]{footnote}%
  \BibitemOpen
  \href@noop {} {}\bibinfo {note} {The magnetization of Li$_2$CuP$_2$O$_7$ is
  marginally increasing above $H_s$. This effect may be ascribed to the
  residual temperature-independent contribution $\chi_0$ and the uncertainty in
  the background compensation at high fields.}\BibitemShut {Stop}%
\end{thebibliography}
%

\end{document}